\documentstyle[12pt,aaspp4,psfig]{article}
\newcommand{\etal}{{\it et al. \,}}

\lefthead{Aparicio \etal}
\righthead{}

\begin{document}
\title{IAC-star: a code for synthetic color-magnitude diagram computation}

\author{Antonio Aparicio} 
\affil{Departamento de Astrof\'\i sica, Universidad de La Laguna/Instituto de
Astrof\'\i sica de Canarias. V\'\i a L\'actea s/n. 
E38200 - La Laguna, Tenerife, Canary Islands, Spain}

\author{Carme Gallart} 
\affil{Instituto de Astrof\'\i sica de Canarias. V\'\i a L\'actea s/n. 
E38200 - La Laguna, Tenerife, Canary Islands, Spain}

\begin{abstract}
  
  The code IAC-star is presented. It generates synthetic HR and
  color-magnitude diagrams (CMDs) and is mainly aimed to star
  formation history studies in nearby galaxies. Composite stellar
  populations are calculated on a star by star basis, by computing the
  luminosity, effective temperature and gravity of each star by direct
  bi-logarithmic interpolation in the metallicity and age grid of a
  library of stellar evolution tracks.  Visual (broad band and HST) and infrared magnitudes
  are also provided for each star after applying bolometric
  corrections. The Padua (Bertelli et al. 1994, Girardi et al. 2000)
  and Teramo (Pietrinferni et al. 2004) stellar evolution libraries
  and various bolometric corrections libraries are used in the current
  version. A variety of star formation rate functions, initial mass
  functions and chemical enrichment laws are allowed and binary stars
  can be computed.  Although the main motivation of the code is the
  computation of synthetic CMDs, it also provides integrated masses,
  luminosities and magnitudes as well as surface brightness
  fluctuation luminosities and magnitudes for the total synthetic
  stellar population, and therefore it can also be used for population
  synthesis research.
  
The code is offered for free use and can be executed at the site {\tt
http://iac-star.iac.es}, with the only requirement of referencing
this paper and crediting as indicated in the site.

\end{abstract}


\section{Introduction \label{intr}}

Galaxies evolve on two main paths: dynamically, including interactions and
merging with external systems and through the process of formation, evolution
and death of stars within them. The latter has the following relevant effects
on the galaxy: (i) the evolution of gas content, (ii) the chemical enrichment
and (iii) the formation of the stellar populations with different properties
as the gas from which they form evolves. The star formation history is
therefore fundamental to understand the galaxy evolution process.

The color-magnitude diagram (CMD) is the best tool to study and derive the
star formation history (SFH) of a galaxy. Deep CMDs display stars born
all over the life-time of the galaxy and are indeed fossil records of the SFH.
A qualitative sketch of the stellar populations present in a
galaxy can be done from a quick look to a good CMD. The presence of stars in
characteristic evolutionary phases indicates that star formation took place
in the system in different epochs of its history. For example, the
presence of RR-Lyre stars is indicative of an old, low metallicity stellar
population; a substantial amount of red-giant branch (RGB) stars is associated
to an intermediate-age to old star formation activity; a well developed
red-tail of asymptotic giant branch (AGB) stars shows that intermediate-age
and young stars with relatively high metallicity are present in the system,
and even a few blue, bright stars as well as HII regions are evidences of a
very recent star formation activity.

A higher degree of sophistication is provided by isochrone fitting to 
significant features of the CMD. Indeed, this method is simple and powerful
enough to determine age and metallicity of single stellar populations as the
ones present in star clusters. 

However, actually deciphering the information contained in a complex CMD and
deriving a quantitative, accurate SFH is complicated and requires more
sophisticated techniques. The standard procedure involves three main
ingredients: (i) good data, from which a deep, ideally reaching the oldest
main-sequence turn-offs, observational CMD can be plotted; (ii) a stellar
evolution library complemented by a bolometric correction library providing
colors and magnitudes of stars as a function of age, mass and metallicity,
and (iii) a method to relate the number of stars populating different regions
of the observational CMD with the density distribution of stars in the CMD as
a function of age, mass and metallicity as predicted by the stellar evolution
theory. The result of this comparison provides the SFH of the system.

Usually, item (iii) involves computation of one or several synthetic CMDs
from which the star density distribution predicted from theory is
deduced. Indeed, the synthetic CMD technique is the most powerful one
available for this kind of analysis and, in general, for the study of the SFH
of resolved stellar populations. The first to compute and use a synthetic CMD
using a monte-carlo technique was Maeder (1974). Chiosi, Bertelli, \& Bressan
(1988) presented the first application of the Padua synthetic CMD code. Tosi
et al.  (1991) used, for the first time, luminosity functions derived from
synthetic CMDs to sketch recent SFHs of nearby galaxies. Bertelli et al.
(1992) were the first to use the global detailed morphology together with
number counts and ratios of star counts in different areas of the synthetic
and observed CMDs to derive the SFH of a galaxy (the LMC).

A more recent version of the Padua code is ZVAR, which distinctive feature is
the age and metallicity interpolation that is performed for each single star
in the Padua grid of stellar evolution tracks. This has been a heavily used
code in the recent past (e.g., Gallart et al. 1996; Vallenari et al. 1996a,b;
Aparicio, Gallart, \& Bertelli 1997a,b; Hurley-Keller, Mateo, \& Nemec 1998;
Gallart et al.  1999; Aparicio, Tikhonov, \& Karachentsev 2000; Vallenari,
Bertelli \& Schmidtobreick 2000; Bertelli \& Nasi 2001; Ng et al. 2002;
Bertelli et al. 2003; see Aparicio et al. 1996 for several examples of
synthetic CMDs using ZVAR and Aparicio 2003, for a review).  Other codes have
been developed by different groups for the computation of synthetic CMDs (see
Tolstoy \& Saha 1996; Dolphin 1997; Holtzman et al. 1999; Ardeberg et al.
1997; Stappers et al. 1997; Dohm-Palmer et al. 1997; Hern\'andez,
Valls-Gabaud, \& Gilmore 1999; Harris \& Zaritsky 2001).

In this paper we present a new algorithm for the computation of synthetic
CMDs. It is intended to be as general as possible. For this purpose, a
variety of inputs for the initial mass function (IMF), SFH, metallicity law
and binariety are allowed. Stars with age and metallicity following a
continuous distribution are computed through interpolation in a stellar
evolution library, providing synthetic CMDs with smooth, realistic stellar
distributions. This significantly improves the accuracy of the comparison
with observed CMDs as compared to with methods without such interpolation.
The Padua stellar evolution libraries by Bertelli et al. (1994; Bertelli94
from here on) and Girardi et al. (2000; Girardi00; from here on) and Teramo
(Pietrinferni et al. 2004) stellar evolution library are used in the current
version of the program, completed by the Cassisi et al. (2000) models for
very low mass stars.  Together with the luminosity and temperature of each
star, the code provides the absolute magnitudes in Johnson-Cousins $UBVRI$,
HST $F218W$, $F336W$, $F439W$m $F450W$, $F555W$ and $F814$ and infrared
$JHKLL'M$ filters. To this purpose the bolometric correction libraries by
Lejeune, Cuisinier, \& Buser (1997), Castelli \& Kurukz (2002), Girardi et
al. (2002) and Origlia \& Leitherer (2000) are used, the later for the HST
filter system. In summary, in the current version of the code, the user can
select among three stellar evolution libraries and four bolometric correction
libraries. The code includes the thermally pulsating AGB (TP-AGB) following
the prescription of Marigo, Bressan, \& Chiosi (1996).

The code is made available for free use. It can be executed at the
internet site {\tt http://iac-star.iac.es} with the only requirement
of crediting as mentioned in that page. The present paper should be
considered as the reference for the code. This paper will be
complemented by several others presenting, in particular, a method
based on a genetic algorithm to compute the SFH from comparison of an
observed and a generic synthetic CMD (Aparicio \& Hidalgo 2004) and a
library of population synthesis based in surface brightness
fluctuations (Mar\'\i n-Franch \& Aparicio 2004).

This paper is organized as follows. In \S\ref{fundcmd}, the fundamentals of
synthetic CMD computation are given, including short descriptions of the
stellar evolution and bolometric correction libraries used. In
\S\ref{stevol}, \S\ref{ftcmd} and \S\ref{cel}, the basic stellar evolution,
the main features arising in a real CMD and the main aspects of chemical
enrichment laws are summarized. These topics are relevant for a better
understanding of the computation process and the synthetic results. Section
\ref{outline} is devoted to explain how the code works as well as some
particularities including the computation of mass loss in advanced
evolutionary phases. Section \ref{run} is a short outline of the code input
and output. A full commented list of input choices and a description of the
output file content are available in the internet site from which the program
is run ({\tt http://iac-star.iac.es}). This will be permanently updated,
including any future change in the input and output.  Finally, a summary of
the paper is given in \S\ref{conc}.

\section{Fundamentals of synthetic CMD \label{fundcmd}}

The distribution of stars in a CMD depends on several aspects on the
physics of stars and galaxies. Broadly speaking, the density of stars
in a given region of the CMD is directly proportional to the time that
stars of any mass and metallicity spend in that region during their
evolution. This information is provided primarily by stellar
evolution theory, on which the computation of synthetic CMDs strongly
relies, complemented by bolometric corrections. But, since the
fraction of stars formed with different masses vary with mass, the
distribution of the stellar masses at birth, the so called IMF,
provides a second, important input characterizing the star density
distribution on the CMD. Stars, on the other hand, are not formed at
the same rate during the life-time of a galaxy. The function
accounting for the star formation rate (SFR) as a function of time is
a further important input. Besides this, stars are born with the
metallicity of the interstellar medium from which they form. So, the
chemical enrichment law (CEL) of the system holding them is another
fundamental information. The last important contribution to the
morphology of the CMD comes from the fraction and mass ratio distribution of
binary stars.  Other physical properties do affect that morphology,
like the internal, differential reddening or the mass distribution (stellar
population gradients, depth effects, etc)
within the system. However, these are properties that can be
considered as external to the CMD itself. In most (or all) cases in
which we are interested, they can be added later on to the computed
synthetic CMD.

In summary, the inputs necessary to compute a synthetic CMD are the
following:

\begin{itemize}
  
\item A stellar evolution library, covering a wide range of stellar
  evolutionary phases and stellar masses and metallicities. It should provide
  stellar luminosities, temperatures and surface gravities as a function of
  mass, metallicity and age.

\item A bolometric correction library to transform from stellar
  luminosities, temperatures and surface gravities to magnitudes in a given
  photometric system. 

\item An IMF, $\phi(m)$.

\item A SFR function, $\psi(t)$.

\item A CEL, $Z(t)$.
  
\item A function accounting for the fraction, and secondary to primary mass
  ratio distribution, of binary stars, $\beta(f,q)$.

\end{itemize}

In the following we will shortly comment on the stellar evolution and
bolometric correction libraries used here and on the computation algorithm.

\subsection{The stellar evolution libraries}

\begin{figure}
  \centerline{\psfig{figure=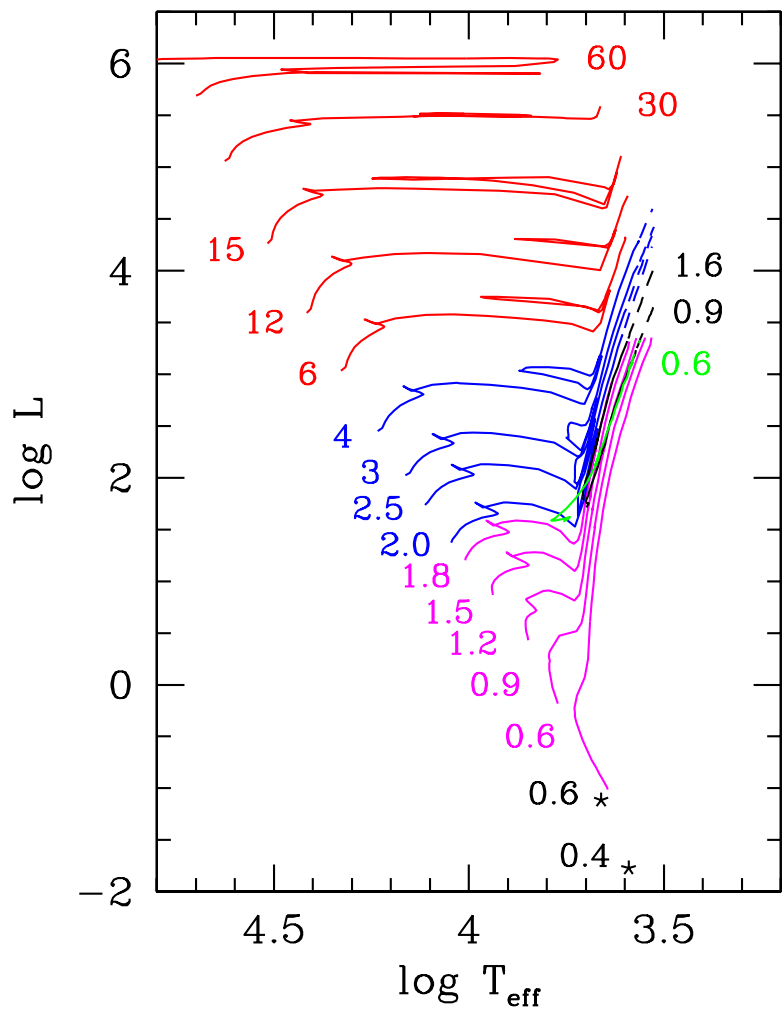,width=16cm}} \figcaption[tracks.eps]{
    Set of evolutionary tracks for metallicity $Z=0.004$ from the library
    of Bertelli94 (full and dashed lines) extended by two
    very low mass models interpolated from Cassisi et al. (2000) of age 10
    Gyr (starred dots). Colors indicate different mass ranges, as follows.
    Red: high mass tracks; blue: intermediate mass tracks; magenta: low mass
    stars up to the He-flash; black: low mass stars HB. In all the cases,
    dashed lines represent the TP-AGB extensions of the Padua tracks computed
    by IAC-star. Labels correspond to the track initial masses, in M$_\odot$.
\label{tracks}}
\end{figure}

Computation of a synthetic CMD relies on a stellar evolutionary library,
which should be computed using the most up-to-date physics, well tested
against observations, and complete in terms of ages, metallicities and
evolutionary phases. These three characteristics are hard to be accomplished
and, indeed, many of the accuracy and precision limitations of synthetic CMD
based techniques originate in this. In its current version, IAC-star uses the
Bertelli94, Girardi00 and Teramo (Pietrinferni et al. 2004) stellar evolution
libraries, completed by the Cassisi et al. (2000) models for very low mass
stars. Even though the Bertelli94 library has been basically superseded by
the Girardi00 one, we have chosen to include both libraries since many SFH
studies in the past have been performed using the Bertelli94 one which may be
chosen for comparative studies. It is our intention to include additional
libraries in the future. In Table \ref{modelos} the characteristics and main
input physics of the three libraries are listed. Figure \ref{tracks} shows
several tracks from the Bertelli94 library as well as two very low mass star
models computed from Cassisi et al. (2000).

\placefigure{tracks}
\placetable{modelos}

\subsection{The bolometric correction libraries}

Comparison of synthetic and observed CMDs require transforming the
first from the {\it theoretical} plane (the $T_{\rm eff}-L$ plane) to
the {\it observational} one (the color-magnitude plane). This is done
by adopting a bolometric correction scale and a color - effective
temperature relation. It is important to note that, while the stellar
evolution library is fundamental for the computation of the synthetic
CMD itself, an equally reliable bolometric correction library is
required if synthetic CMDs are to be compared with observed ones. This also
implies that the derivation of accurate SFHs relies on the accuracy of
both stellar evolution library and the model atmospheres set adopted
for computing the bolometric correction scale and color - $T_{\rm eff}$
relation.

In its current version, IAC-star implements four choices for the
bolometric correction and color - $T_{\rm eff}$ relation libraries:

\begin{itemize}
\item The Girardi et al. (2002) library providing bolometric corrections for
  visual, broad band $UBVRI$ and infrared $JHK$ filters.
\item The Castelli \& Kurucz (2002) library for visual, broad band $UBVRI$
  and infrared $JHKL$ filters, completed with the Flucks et al. (1994)
  semi-empirical transformations for M giants. Care has been taken that the
  two sets properly match.
\item The Lejeune, Cuisinier, \& Buser (1997) library for visual, broad band
  $UBVRI$ filters and infrared $JHKLL'M$ filters.
\item The Origlia \& Leitherer (2000) library for $F218W$, $F336W$, $F439W$,
  $F450W$, $F555W$ and $F814W$ filters of the HST WFPC2 and ACS.
\end{itemize}

These libraries fulfill the requirements of completeness and wide
metallicity ($[Fe/H]$), temperature ($T_{\rm eff}$) and gravity ($g$)
coverage and sampling, as well as having been empirically
calibrated. It is our intention to include other libraries, as well as
transformations for additional photometric bands such as the Stromgren
ones, in the future as they become available.

The bolometric correction and the color - $T_{\rm eff}$ are functions of the
following three parameters: $BC([Fe/H],T_{\rm eff},g)$. The used libraries
allow obtaining bolometric corrections for almost all cases, although they
fail to fully cover a few, very low temperature cases. Linear extrapolation
is used in such cases. Since it is necessary in very few cases, the deviation
effects that could arise in the CMD from that extrapolation should be
negligible. However, detailed analysis of color distribution of stars in
phases as the extended AGB should be made with caution.

\subsection{Computing a synthetic CMD}

IAC-star uses a monte-carlo method to compute synthetic CMDs on a star by
star basis. A random number generator (RNG) is used to compute, first the
mass, and then the time of birth of the star, according to the distribution
functions $\phi(m)$ and $\psi(t)$, respectively.  The value of time is
introduced in the function $Z(t)$ to obtain the metallicity. If metallicity
dispersion is allowed, its exact value is obtained using such a dispersion
and a new random number. The binariety of the star is determined in a similar
way, using a RNG and the function $\beta(f,q)$, in which $f$ is the fraction
of binaries and $q$ the internal mass rate. If the star turns out to be a
binary, it is assumed to be the primary (most massive) star of the system.
The age and metallicity of the secondary are assumed to be the same as for
the primary. The mass of the secondary is computed using again the RNG and
according to the secondary to primary mass ratio distribution provided by $q$
and an assumed IMF for the secondaries. The lifetime of each star, according
to the stellar evolution models, is used at this point to decide whether the
star is alive (and to calculate then its remaining parameters), or dead (and
to calculate the properties of the remnant). The case in which only the
secondary star is alive is allowed.

Once the mass, age and metallicity are known, luminosity, temperature and
surface gravity of the star are computed through interpolation within the chosen
stellar evolutionary library. If the star is a binary, the process is repeated
for each component. Finally, luminosity, temperature and gravity are
interpolated in the chosen bolometric correction library to obtain magnitudes and
colors in the standard photometric system.

These are the computation fundamentals. In \S\ref{run}, we provide
details about the procedure itself. But before, it is clarifying to
outline the aspects of the stellar evolution relevant to the
morphology of the CMD (\S\ref{stevol}); to review such morphology
(\S\ref{ftcmd}), and to discuss the CEL, its parameterization and its
relation to basic physical properties of the system (\S\ref{cel}).

\section{Basic stellar evolution \label{stevol}}

\begin{figure}
  \centerline{\psfig{figure=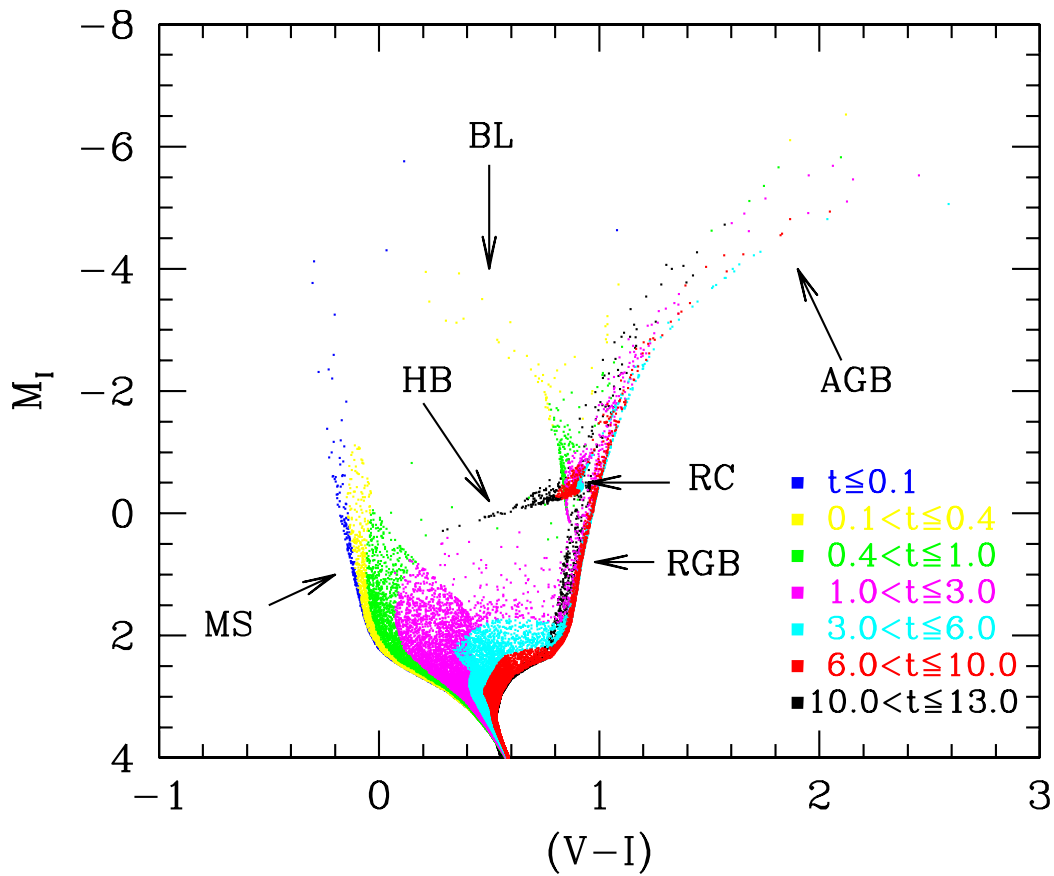,width=16cm}}
  \figcaption[sin1.eps]{Synthetic CMD computed using constant SFR from 13 Gyr
    ago to date and metallicity linearly increasing from $Z_0=0.0001$ to
    $Z_f=0.02$ and a Kroupa et al. (1993) IMF. The Bertelli94 stellar
    evolution library and the Lejeune et al. (1997) bolometric correction
    library have been used. Stars in different age
    intervals are plotted in different colors. The color code is given in the
    figure, in Gyr. Labels indicate the evolutionary phase (see text for an
    explanation).
\label{sin1}}
\end{figure}

Luminosity, $L$, and temperature $T_{\rm eff}$, as a function of
stellar age and metallicity, as well as lifetime, are, for our
purposes, the main information provided by the stellar evolutionary
library. If $s$ is a star of mass $m_{\rm s}$, metallicity $Z_{\rm s}$
and age $a_{\rm s}$, a first estimate of its location in the CMD might
be obtained just by time interpolation within a single evolutionary
track of age and metallicity closest to the stellar values. However,
if a realistic synthetic CMD is to be computed, interpolation must be
done between tracks both in age and metallicity. The fact that stars
experience different particular physical processes along their
evolution that, like the He-flash, abruptly interrupt their smooth
evolutionary paths, makes this interpolation rather cumbersome. In the
following, we will sketch the basic stellar evolution directly
affecting the synthetic CMD computation problem. In this sense, it is
useful to divide stars in three groups according to their masses: low,
intermediate and high mass stars (see Chiosi et al. 1992 and Chiosi
1998 for a short but clear summary of the basic stellar evolution).
The critical masses separating these groups depend on the chemical
composition and the input physics. But, roughly, the mass separating
the low and intermediate dominions, $m_{\rm lw}$, is in the range
$1.7-2.0$ M$_\odot$ and the mass separating the intermediate and high
dominions, $m_{\rm up}$, is in the range $5-7$ M$_\odot$. In
Fig. \ref{sin1} the loci of the different features that
will be introduced in the following are displayed.

Stars of any mass spend a major fraction of their lives in the core
H-burning, main sequence (MS) phase. As H is exhausted in the core, the star
leaves the MS while the H-burning proceeds in a shell. At this point, a star
may follow two types of paths in the CMD, depending on its mass. In low mass
stars, the H-exhausted, He core is not massive, hot and dense enough to
immediately start the He-burning. The core gravitationally contracts and, as
a consequence, degeneracy appears in the electronic gas. This degeneracy
stops the core collapse. As H-burning proceeds in a shell, the star climbs
the RGB and more He is incorporated into the degenerated core.  When the mass
of this core reaches about 0.45 to 0.50 M$_\odot$ (the precise mass depends
on chemical composition, stellar mass and input physics), He-burning starts
violently, removing the electron degeneracy. This is the so called He-flash.
The star abruptly terminates the RGB phase and start the horizontal-branch,
in which He-burning and H-burning proceed in the core and a shell,
respectively. Since the He-flash is produced at essentially identical core
mass, the maximum RGB luminosity is almost constant, independently of the
stellar metallicity and initial mass. The same happens to the luminosity of
the horizontal branch (HB). However, the stellar surface effective
temperature, which determines the position of the star in the HB, strongly
depends on several factors, among which the most important are the chemical
composition and total stellar mass. It must be noted that the latter depends
not only on the initial stellar mass, but also on the mass loss during the
RGB and the He-flash.

In intermediate and high mass stars, the He-core ignites in non-degenerate
conditions as soon as the central temperature and density reach some
threshold values, approximately $10^8$ K and $10^4$ g cm$^{-3}$,
respectively. This requires a minimum core mass of 0.33 M$_\odot$. Since this
mass depends on the initial stellar mass, the luminosity of the star during
this phase increases with stellar mass. In intermediate mass stars (including
the lower end of high mass stars), the core He-burning phase takes place in
two regions of the CMD. The first one is close to the Hayashi line.
Subsequently, when the energy produced in the He core, which is increasing,
equals the energy produced in the H shell, which is decreasing, the outer
envelope rapidly contracts passing from convective to radiative. Temperature
increases and the star migrates to a bluer region. This produces the
blue-loop (BL) phase. It appears rather red in Fig. \ref{sin1} as corresponds
to the high metallicity of the computed young population.

Both in low and intermediate mass stars, a C-O core is formed as a
result of core He-burning. Upon the exhaustion of He in the core, the
mass of the C-O core is constant for low mass stars, but increases
with stellar mass for all the others. The structure of stars is hence
formed by a C-O core, a He-burning shell, a H-burning shell and a
H-rich envelope. In low and intermediate mass stars, the mass of the
C-O core is not high enough to ignite and an electron degenerated core
again develops. The star expands, produces a convective envelope and
evolves upward running almost parallel to the RGB. This is the
asymptotic giant branch (AGB) phase. Cooling in the regions external
to the C-O core extinguishes the H-burning shell but, eventually, the
expansion of the envelope is stopped by its own cooling, the envelope
contracts and the H-shell re-ignites. This terminates the early-AGB
(E-AGB). After that, the He-shell becomes thermally unstable. Both,
the H and He shells alternate as the main sources of energy producing
the TP-AGB phase. This process results in efficient mass loss and
terminates with the complete ejection of the envelope and temporally
producing a planetary nebula. The remnant is a C-O white dwarf of mass
lower than the Chandrasekar limit (1.4 M$_\odot$). Intermediate (and
low) mass stars fail to ignite the C-O core.

In high mass stars, the C-O core ignites in non-degenerate conditions. The
fraction of high mass stars formed in a galaxy is small and their life-times
short. For these reasons they do not affect significantly the structure of
the CMD and we do not go further in the discussion of their evolution. It is
enough to mention that mass loss plays a fundamental role in determining the
position in the CMD of core He-burning stars more massive than about 15
M$_\odot$, and that, after core C-O ignition, they proceed through a series
of nuclear burnings and finish theirs lives with a supernova explosion.

\section{Main features of the CMD \label{ftcmd}}

The main features and the distribution of stars produced by stellar evolution
in a CMD can be seen in Fig. \ref{sin1}, where a synthetic CMD, computed for
constant SFR from 13 Gyr ago to date, is shown. The CMD is based in the
Bertelli94 stellar evolution library and the Lejeune et al. (1997) bolometric
correction library. The IMF obtained by Kroupa, Tout \& Gilmore (1993) and a
simple CEL with metallicity Z(t) increasing linearly with time, initial value
$Z_0=0.0001$ and final value $Z_{\rm f}=0.02$ have been used. Colors are
associated to age intervals to illustrate the distribution of stars in the
CMD according to their age. Labels are included to identify the MS, BL, RGB,
AGB, HB and red-clump (RC) phases. The RC is formed by a mix of the reddest
HB stars and the bottom part of the core He-burning phase of intermediate
mass stars. The remaining features have been introduced in \S\ref{stevol}.

The MS provides the most important and unambiguous age information. But it is
interesting to note that, besides it, the BL and the AGB, if well populated,
provide useful and complementary age information, which is poor, however, on
the RGB. But the latter gives information on the integrated star formation
rate for intermediate and old stellar populations and provides clues to
constrain the CEL.

\placefigure{sin1}

\begin{figure}
  \centerline{\psfig{figure=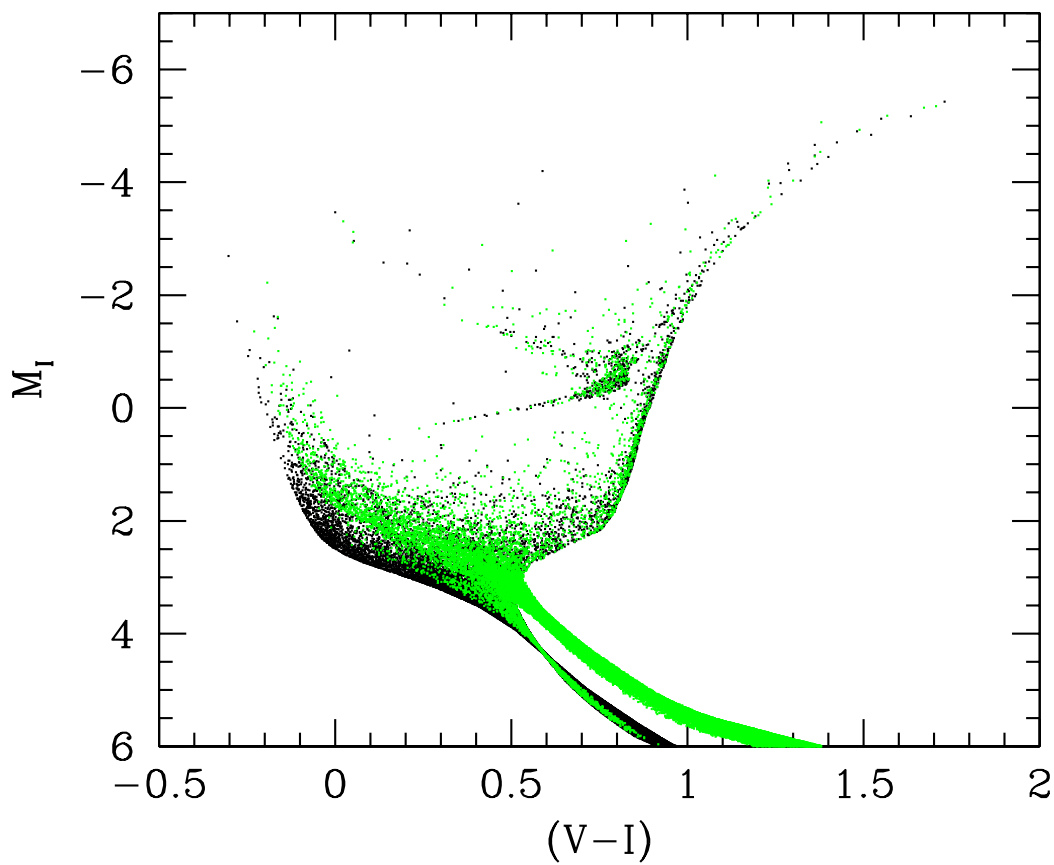,width=16cm}}
  \figcaption[sin1.eps]{Synthetic CMD computed using constant SFR from
  13 Gyr ago to date and metallicity linearly increasing from
  $Z_0=0.0001$ to $Z_f=0.001$, Salpeter IMF, and a 0.3 fraction of
  binary stars with minimum mass ratio 0.7. The Bertelli94 stellar
    evolution library and the Lejeune et al. (1997) bolometric correction
    library have been used. Colors and magnitudes of
  binary stars are represented in green. Green points in the single
  star locus of the MS correspond to binary stars in which the primary is
  dead.
\label{binz0z001}}
\end{figure}

The morphology of the CMD is altered by the presence of binary stars. 
Figure~\ref{binz0z001} illustrates the effect, which is most obvious in the
lowest MS but affects also the overall distribution of stars in the CMD.

To simplify the discussion of stellar populations based on CMD
analysis, it is useful to introduce some terminology. We will use the
terms old, intermediate-age and young to specify three stellar age
ranges which are associated to features clearly recognizable in the
CMD. By {\it old} we will refer to populations that are not able to
produce bright, extended AGBs. This corresponds to ages larger than
about 10 Gyr. Good examples of old objects are globular clusters. Old
populations will always produce a well populated RGB and additionally,
they may show a blue extended HB. However, the latter is not a
unambiguous age indicator, since it depends on metallicity and mass
loss on the RGB. {\it Intermediate-age} populations are those
producing an RGB as well as extended, bright AGBs. They are older than
$1-2$ Gyr. Finally, {\it young} populations are those igniting He in
the nucleus in non-degenerate conditions, and therefore failing to
produce an RGB. The age separating the young from the intermediate-age
ranges is the MS life time of stars of mass equal to $m_{\rm lw}$, the
mass separating the low and intermediate-mass dominions. A pure young
population would not have had time for low-mass stars to evolve from
the MS.

\section{The chemical enrichment law of a stellar system \label{cel}}

In its current version, IAC-star offers two main procedures to compute the
CEL, $Z(t)$: (i) by linear interpolation in several, arbitrary $Z(t)$ nodes
given by the user, and (ii) from the parameters (yield, infall and outflow
rates, etc.) used in physical CEL scenarios. For a better understanding of
the second procedure, the CEL fundamentals are outlined in this section.
However, it must be noted and kept in mind that IAC-star is not a code to
solve for chemical enrichment law scenarios. It allows a wide range of
choices for them and computes stars with the necessary coupling between age,
metallicity, SFR and gas fraction provided by the input parameters. But it is
the user responsibility to decide what kind of chemical law is to be used. On
the other hand, the linear interpolation in $Z(t)$ nodes, which includes a
linear law and a constant one, is a simplistic, perhaps physically
unrealistic choice, but it may work well for many users, in particular when
details about the actual metallicity law are not known. It must be also noted
that a simple, linear CEL is not far from the law of a moderate infall model.
  
Metallicity laws for a closed box and for infall, outflow of well mixed
material and outflow of rich material can be found in Peimbert, Col\'\i n, \&
Sarmiento (1994). In IAC-star the infall and outflow (of well mixed material)
scenarios can be used as well as the close box one, which is but a limit case
of any of the former. The unmixed, rich material outflow can be simulated by
the well mixed material outflow if the yield is allowed to vary and we are
not interested on the Helium to Oxygen ratio, which, in any case, is beyond
the scope of this software.

Assuming instantaneous recycling, infall and outflow laws are given by
(Peimbert et al. 1994)

\begin{equation} \label{infall}
Z(t)=Z_0+\frac{y}{\alpha}\{1-[\alpha-(\alpha-1)\mu(t)^{-1}]^{-\alpha/(1-\alpha)}\}{\hskip 1.5cm}{\rm (Infall)}
\end{equation}
\begin{equation} \label{outflow}
Z(t)=Z_0+\frac{y}{\lambda+1}\ln[(\lambda+1)\mu(t)^{-1}-\lambda] {\hskip 1.5cm}{\rm (Outflow)}
\end{equation}

The chemical enrichment theory has been developed elsewhere (see
Tinsley 1980; Peimbert et al. 1994 and references therein) and it is
not our aim to discuss it here in any detail. However, the former
equations deserve some comments in order to understand the CEL
computation in IAC-star. The initial metallicity, $Z_0$ should be 0
for physically realistic models, but it will be useful to have the
possibility of assuming some initial enrichment in some cases. The
yield, $y$ is defined as the mass of newly formed metals that a
generation of stars ejects to the interstellar medium, relative to the
mass locked in stars and stellar remnants by the same stellar
generation.  $\mu$ is the gas mass fraction relative to the total mass
intervening in the chemical evolution; i.e.  including stars, stellar
remnants, low mass objects, dust and gas itself but excluding dark
matter, which does not intervene in such process.

Parameters $\alpha$ and $\lambda$ control infall and outflow respectively.
They are defined as $f_{\rm I}=\alpha(1-R)\psi$ and $f_{\rm
  O}=\alpha(1-R)\psi$, where $f_{\rm I}$ and $f_{\rm O}$ are the infall and
outflow rates, respectively and $R$ is the mass fraction returned to the
interstellar medium by a generation of stars.  In this way, $(1-R)\psi$ is
the mass locked into stars and stellar remnants and $f_{\rm I}$ and $f_{\rm
  O}$ are given as fractions ($\alpha$ and $\lambda$) of this mass. A few
things must be pointed out. First, since $\psi$ is a function of time, also
$f_{\rm I}$ and $f_{\rm O}$ are so. Second, both $\alpha$ and $\lambda$ can
take values $>1$. Finally, the simple closed box model is obtained from any
of the former relations if, respectively, $\alpha=0$ or $\lambda=0$.

The mass fraction, $\mu$, appearing in equations \ref{infall} and
\ref{outflow} is coupled to the SFR or, rather, to its integral, by

\begin{equation} \label{mu}
\mu(t)=1-\frac{(1-R)\Psi(t)}{M_0+(1-R)\Psi(t)(\alpha-\lambda)}
\end{equation}

\noindent where $M_0$ is the initial mass of the system and $\Psi(t)=\int_0^t\psi(t'){\rm d}t'$. This
relation for $\mu$ is valid for equation \ref{infall} and \ref{outflow} just
using $\lambda=0$ in the first case and $\alpha=0$ in the second one. Lets
assume for simplicity, and only in the context of the chemical law
computation, that the time unit is the present age of the system, with $t=0$
for the initial instant and $t=1$ for the present day time, and that the
integral of the SFR is normalized also for the system age
($\int_0^1\psi(t'){\rm d}t'=1$). With this, the initial mass of the system
can be written as

\begin{equation}\label{mo}
M_0=(1-R)(\lambda-\alpha-\frac{1}{\mu_{\rm f}-1})
\end{equation}

\noindent where $\mu_{\rm f}$ is the final gas fraction. The yield, which is
assumed here to be constant, can also be expressed as a
function of $Z_0$ and $Z_{\rm f}$ (the final metallicity):

\begin{equation}\label{yin}
y=\frac{(Z_{\rm f}-Z_0)\alpha}{1-[\alpha-(\alpha-1)\mu_{\rm
    f}^{-1}]^{-\alpha/(1-\alpha)}} {\hskip 1.5cm}{\rm (Infall)}
\end{equation}
\begin{equation}\label{yout}
y=\frac{(Z_{\rm f}-Z_0)(\lambda+1)}{\ln[(\lambda+1)\mu_{\rm f}^{-1}-\lambda]} 
{\hskip 1.5cm}{\rm (Outflow)}
\end{equation}

In practice, the most frequent situation is one in which the
researcher has observational information about $\mu_{\rm f}$ and
$Z_{\rm f}$ and can assume that $\mu_0=1$ and $Z_0=0$ or small. $R$
can be derived from the stellar evolution models (it is $R\simeq 0.2$
for the Padua stellar evolution models, although it depends slightly
on the metallicity) and $\psi(t)$ is the key function of the stellar
population model (usually what is sought in the analyzed galaxy). The
remaining parameters are $\alpha$ and $\lambda$, for which reasonable
guesses need to be made. Of course, in this way $y$ is no longer the
physical stellar yield, but an {\it effective} yield defined just by
equations \ref{yin} and \ref{yout}. If the physical relevance of the
chemical evolution model is seeked, the consistency of all the
parameters must be analyzed. However, for the purposes of computing a
synthetic CMD, the chemical law is only an ingredient of the code,
which can be imposed from outside. Users of IAC-star are encouraged to
check and decide the required chemical law. We insist that IAC-star
only provides the coupling between $\psi(t)$, $\mu(t)$ and Z(t) for
the set of input $Z_0$, $Z_{\rm f}$, $\mu_{\rm f}$ and $\alpha$ or
$\lambda$.

\begin{figure}
  \centerline{\psfig{figure=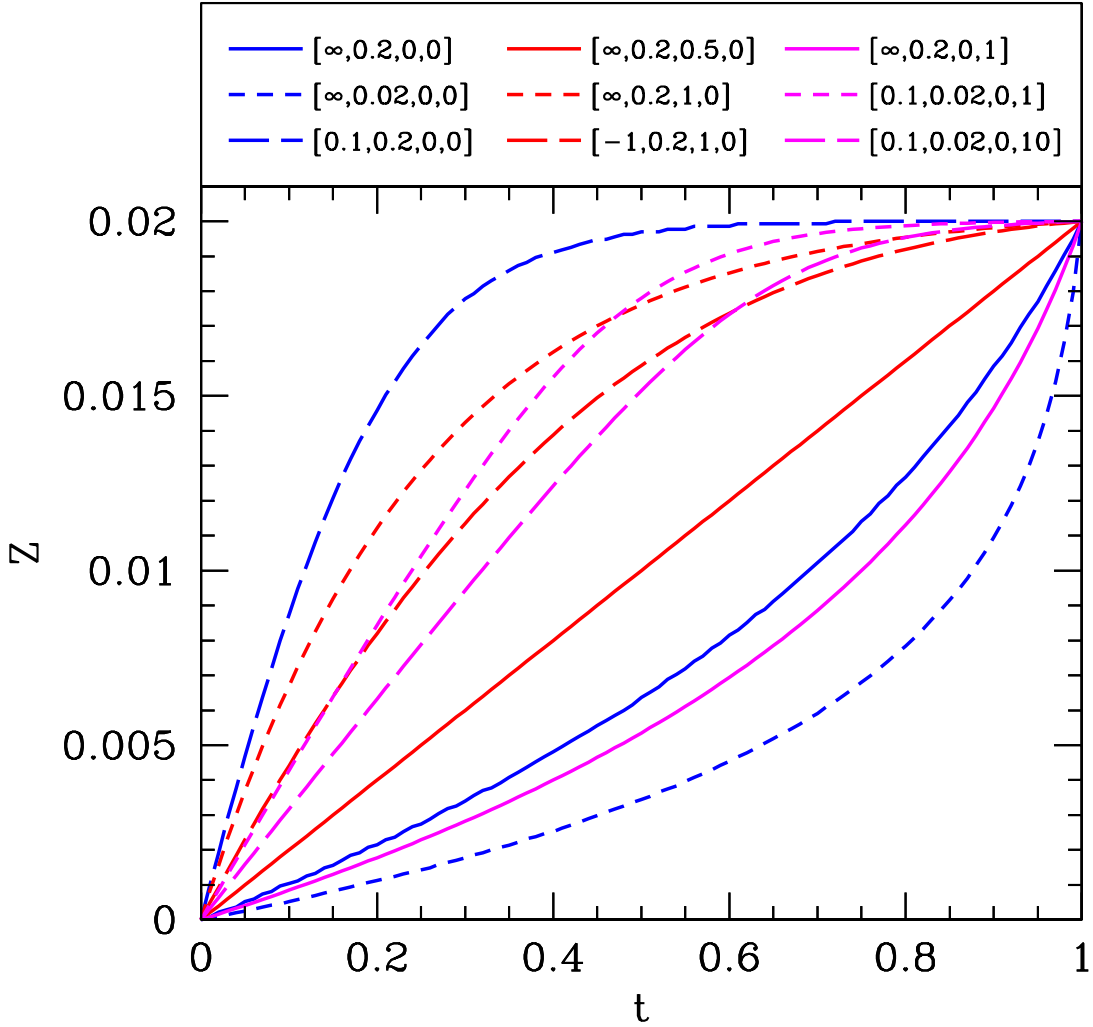,width=16cm}}
  \figcaption[leyz1.eps]{ Examples of CELs for different choices of
  $\mu_{\rm f}$, $\alpha$, $\lambda$ and $\psi(t)$ are given for fixed
  $Z_0=0$ and $Z_{\rm f}=0.02$.  For $\psi(t)$, an exponential law of
  the form $\psi(t)=A\,{\rm exp}(-t/\beta_\psi)$ has been used, where
  $A$ is set to normalyze the integral of $\psi(t)$ for the entyre
  life of the systen. Time scale is normalyzed, with the instant $t=0$
  corresponding to the beginning of the star formation and the instant
  $t=1$ corresponding to its end. In the upper window the parameter
  choices are displayed. Bracketed figures correspond to
  $[\beta_\psi,\mu_{\rm f},\alpha,\lambda]$. Closed-box models are
  drawn in blue, infall models are drawn in red and outflow models in
  magenta.
\label{leyz1}}
\end{figure}

\begin{figure}
  \centerline{\psfig{figure=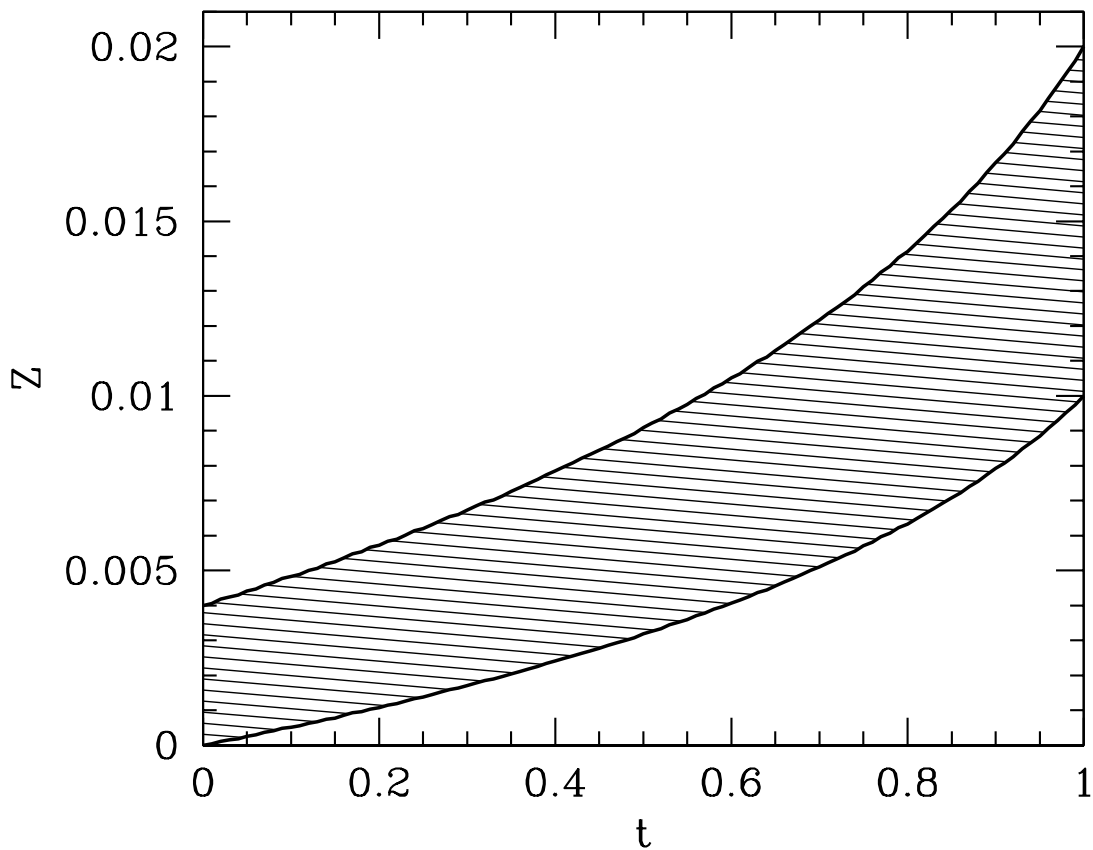,width=16cm}}
  \figcaption[leyz2.eps]{ Example of a CEL with metallicity
  dispersion. A constant SFR has been used with $\alpha=\lambda=0$ and
  $\mu_{\rm f}=0.2$ in both cases, but $Z_0=0$, $Z_{\rm f}=0.01$ for
  the lower law and $Z_0=0.004$, $Z_{\rm f}=0.02$ for the upper
  one. Time scale is normalyzed, with the instant $t=0$ corresponding
  to the beginning of the star formation and the instant $t=1$
  corresponding to its end. For each age, stars would have
  metallicities comprised within the shaded area.
\label{leyz2}}
\end{figure}

\placefigure{leyz1}
\placefigure{leyz2}

In Fig. \ref{leyz1} several examples of CELs for different choices of
$\mu_{\rm f}$, $\alpha$, $\lambda$ and $\psi(t)$ are given for fixed
$Z_0$ and $Z_{\rm f}$. For $\psi(t)$, an exponential law of the form
$\psi(t)=A\,{\rm exp}(-t/\beta_\psi)$ is used. In Fig. \ref{leyz2} a law with
metallicity dispersion is shown. 

\section{IAC-star computation outline \label{outline}}

In this section we will outline how the IAC-star algorithm works paying
attention to the general interpolation procedure and to how different stellar
mass ranges and evolutionary phases are handled. Computation of mass loss in
advanced evolutionary phases is explained in the last paragraph. In
Fig. \ref{tracks} several tracks of different masses from the Bertelli94
library are plotted as reference
for the following discussion.

\subsection{The general case}

As discussed in \S\ref{fundcmd}, a monte-carlo algorithm and the corresponding probability
distribution functions are used to provide the age, $a_{\rm s}$, mass,
$m_{\rm s}$, and metallicity, $Z_{\rm s}$, of each synthetic star. In the
general, simplest case, mass and metallicity are used to compute the
evolutionary track of the star by bi-logarithmic interpolation within the
evolutionary track library. It is important to make this simple but
important and cumbersome process clear. Note that one of the difficulties is
that tracks are not functions, but arbitrary curves, time being the driving
parameter. For this, interpolation must be done between homologous sections
of different tracks. 

Let first introduce some notation. Tracks are identified by mass and
metallicity. We will name tracks by $K_{\rm i,j}$, the first index
denoting the metallicity and the second one, the mass. $Z_{\rm s}$
represents a value of metallicity between two values present in the
stellar evolution library. The algorithm input does not allow
extrapolation outside the library limits in Z. Lets call $Z^{\rm
lib}_{\rm i}$ and $Z^{\rm lib}_{\rm i+1}$ these two values,
respectively. So $Z^{\rm lib}_{\rm i}\leq Z_{\rm s}\leq Z^{\rm
lib}_{\rm i+1}$. A search is then made within $Z^{\rm lib}_{\rm i}$
and $Z^{\rm lib}_{\rm i+1}$ track subsets to look for the mass values
embracing $m_{\rm s}$. Lets call $m^{\rm
lib}_{\rm i,j}$ and $m^{\rm lib}_{\rm i,j+1}$ the track mass values
embracing $m_{\rm s}$ for the $Z^{\rm lib}_{\rm i}$ subset and $m^{\rm
lib}_{\rm i+1,k}$ and $m^{\rm lib}_{\rm i+1,k+1}$ the same for the
$Z^{\rm lib}_{\rm i+1}$ subset. So $m^{\rm lib}_{\rm i,j}\leq m_{\rm
s}\leq m^{\rm lib}_{\rm i,j+1}$ and $m^{\rm lib}_{\rm i+1,k}\leq
m_{\rm s}\leq m^{\rm lib}_{\rm i+1,k+1}$. The track corresponding to
star $s$ can now be obtained by logarithmic interpolation between
tracks $K_{\rm i,j}$, $K_{\rm i,j+1}$, $K_{\rm i+1,k}$, $K_{\rm
i+1,k+1}$. Lets call this interpolated track $K_{\rm
s}$. Additionally, lets call $\tau_{\rm i,j}$ the life time of track
$K_{\rm i,j}$, i.e., the time at which evolution within that track
finish, and $\tau_{\rm s}$ the corresponding life time for the
interpolated track.

Once $K_{\rm s}$ has been obtained, the photometric parameters of the
star are obtained straightforwardly by logarithmic interpolation of
$a_{\rm s}$ within this track. However, before this, whether the star
is death or alive must be established. This is done just comparing
$a_s$ with $\tau_{\rm s}$. If $a_{\rm s}\leq \tau_{\rm s}$,
logarithmic interpolation is performed and the computation is finished
for this star. If not, a new dichotomy arises. If the star is of
intermediate or high mass, $a_{\rm s}>\tau_{\rm s}$ means that it is
death. The mass of the remnant is then computed as explained below and
added to the total mass of the system. If the star is of low mass,
$a_{\rm s}>\tau_{\rm s}$ means that it is beyond the
He-flash. Computation in the HB track subset must be performed in this
case.  For this, the mass loss of the star during the RGB is
subtracted from $m_{\rm s}$ (see below) and $\tau_{\rm s}$ is
subtracted from $a_{\rm s}$. The resulting mass and age, $m_{\rm
s,HB}$ and $a_{\rm s,HB}$, are used in the HB track
subset. Computation is similar to that in the general library. If
$a_{\rm s,HB}$ is larger than the life time of the new interpolated HB
track, the star is considered death and the remnant mass is computed
and stored as before.

We have outlined the general computation. However, a few special cases can
occur that need more comments. We will treat them in the following.

\subsection{Interpolation and extrapolation between different mass ranges}

It may occur that the tracks embracing the mass and metallicity of the
synthetic star belong to two different mass ranges (see Fig. \ref{tracks}). If
some tracks are in the high mass range and some in the intermediate mass
range, interpolation can be performed as in the general case up to the end of
the tracks. After that, high mass stars are considered death while
intermediate mass stars continue through the TP-AGB, computed as outlined
below.

The case is somewhat more complicated if some tracks are in the
intermediate-mass range and some in the low-mass range. The stellar physical
processes are qualitatively similar for both ranges up to the basis of the
RGB: core H-burning (MS) or He-core gravitationally contracting plus shell
H-burning (sub-giant phase). Interpolation can hence be performed up to that
point but not afterward, when intermediate mass stars are about to start the
core He-burning while low mass stars evolve climbing the RGB. In this case,
the mass separating the intermediate and low mass ranges is firstly
determined by logarithmic interpolation between the four tracks embracing the
mass and metallicity of the synthetic star. This separating mass is used to
determine the range to which the star belongs. After this, interpolation is
performed up to the RGB basis. Beyond that point, extrapolation is made from
tracks of the mass range to which the star belongs. The CMD related features
can be seen in Fig. \ref{tracks}
  
\subsection{Very high and very low mass stars \label{vlm}}

The stellar mass interval is controled by the input IMF. Stars of masses
beyond the interval covered by the chosen stellar evolution library are
handled in different ways, depending on the mass being larger than the
maximum mass of the library or smaller than the minimum one. In the first
case the star is simply considered death. The case of very low mass stars 
is more complex. 

The smallest mass computed in the used libraries are given in Table
\ref{modelos}. Stars less massive than that have
MS life times larger than the age of globular clusters and are
certainly in the MS for any realistic case.  They are normally below
the limiting magnitude in most CMDs of galaxies. However, they have a
quite significant contribution to the total mass of the system and,
also to the integrated luminosity, magnitudes and colors. Although
this contribution can be estimated by integration of the IMF, it may
be useful to have the possibility of explicitly computing their colors
and magnitudes. To this purpose, the libraries mentioned above
have been completed downward in mass using the models by Cassisi et
al. (2000). They present MS $L$ and $T_{\rm eff}$ of stars with masses
between $\sim 0.09$ M$_\odot$ and 0.8 M$_\odot$ for metallicity values in the 
range: $0.0002\leq 0.002$. For present purpose these models have been
completed with a very low mass sequence for solar chemical composition
(Cassisi 2003, private communication). Since these stars have long
evolution times and the analysis of the CMD does not usually rely on
their exact distribution in it, the effect of age on their $L$ and
$T_{\rm eff}$ can be neglected. A polynomic fit has been performed,
for each metallicity, to the $m-L$ and the $m-T_{\rm eff}$ plots of
the model stars. This can be used to estimate $L$ and $T_{\rm eff}$ of
the synthetic star. The coefficients of the fit for the general
equation

\begin{equation}
y=a_0+a_1x+a_2x^2+a_3x^3+a_4x^4+a_5x^5
\end{equation}

\noindent are given in Table \ref{muybaja}. Here, $x$ stands for
$\log m$ and $y$ for $\log L$ or $\log T_{\rm eff}$, respectively. 

Figure \ref{tracks} shows two models computed with these relations for masses
0.4 and 0.6 M$_\odot$ and $Z=0.004$. A good agreement exists with the
Bertelli94 tracks. Consistence is also quite good comparing with the
Girardi00 and the Teramo libraries (not shown here). However the fact that
the very low mass stars are computed from a different model set (namely the
Cassisi et al. 2000 one) and that no age computation is made for them must be
taken into account if accurate analysis of CMD is intended in the region 
around 0.6 M$_\odot$ for the Bertelli94 and the Teramo libraries and 0.15
M$_\odot$ for the Girardi00 library; i.e., around the lower mass limits of these
libraries.

\placetable{muybaja}

\subsection{Mass loss}

Mass loss by stellar winds plays an important role in the evolution of
massive stars. It is included in the Bertelli94 computation for stars with
initial mass larger than 12 M$_\odot$. The reader is referred
to Bertelli94 for details. Besides massive stars, mass loss during the RGB
and the AGB has dramatic consequences on the evolution of low and
intermediate mass stars. In these cases the effects of mass loss are included
at the moment of computing synthetic stars in IAC-star. Unfortunately, mass
loss during these phases and, in particular, during the AGB depends on poorly
known physical parameters that should be fine-tuned.  In the modeling of
mass loss during these phases we follow the criteria given by Bertelli94 and
Marigo et al. (1996).

Mass loss during the RGB does not affect the internal structure of the
star and can be neglected in the computation of $L$ and $T_{\rm
eff}$. However, the integrated mass loss during this phase determines
the mass envelope of the subsequent HB stars and hence its $T_{\rm
eff}$. For this reason, mass loss can be integrated during the RGB and
introduced in a single step at the tip of the RGB (TRGB).  To this
purpose, the empirical relation by Reimers (1975) is used:

\begin{equation}
\dot{m}=1.27\times10^{-5}\eta m^{-1}L^{1.5}T_{\rm eff}^{-2} {\hskip 3mm}{\rm
  M}_\odot/{\rm yr}
\end{equation}

\noindent where $L$ and $m$ are in solar units and $\eta$ is a scaling
parameter. The lowest stellar initial mass which evolution time up to the
TRGB is smaller than or equal to the globular cluster age (about 13 Gyr) is
about $0.80 - 0.95$ M$_\odot$, depending on the metallicity, lower masses
corresponding to lower metallicities. From this, the mass loss during the RGB
must be subtracted. Since the core mass at the moment of the He-Flash is
about 0.55 M$_\odot$, the envelope mass of these stars can be very small. As
a consequence, the exact value adopted for $\eta$ strongly affects $T_{\rm
eff}$ at the zero-age HB. In the current version of IAC-star, $\eta$ has a
default value of 0.35, but it can be modified by the user. 

Computing the mass loss during the AGB is more complicated. It can be
neglected during the E-AGB, but it is the key parameter controlling the final
evolution of the star during the TP-AGB. Stellar evolution models of low and
intermediate-mass stars are in general computed up to the beginning of the
TP-AGB only. From that point on, computation is performed within IAC-star
following the prescriptions by Marigo et al. (1996). First, the starting
point of the TP-AGB is assumed to be the point in the track just before the
first significant He-shell flash. Hence, the evolution through the TP-AGB is
followed using several basic relations: (i) the mass loss rate; (ii) a
relation connecting the core mass with the total luminosity of the star;
(iii) the rate at which the core mass increases as a result of shell
H-burning, and (iv) an $L-T_{\rm eff}$ relation for the TP-AGB. Details on
these relations can be found in Marigo et al. (1996) and references therein.
We shortly summarize them here for self-consistency.

The mass loss rate is taken as the minimum of the two following values:

\begin{equation} \label{mdot1}
\log\dot m=-11.4+0.0123P
\end {equation}
\begin{equation} \label{mdot2}
\dot m=6.07023\times 10^{-3} L c^{-1} v_{\rm exp}^{-1}
\end {equation}

\noindent where $m$ and $L$ are the stellar mass and luminosity (in
solar units), $P$ is the pulsation period (in days) $c$ is the light speed
(in km s$^{-1}$) and $v_{\rm exp}$ is the terminal velocity of the stellar wind
(in km s$^{-1}$). Equations \ref{mdot1} and \ref{mdot2} stand for periods
shorter and larger than about 500 days, respectively. The wind expansion velocity is
calculated in terms of the period as (Vassiliadis \& Wood 1993):

\begin{equation}  
v_{\rm exp}=-13.5+0.056P
\end {equation}

\noindent and the period is derived from the period-mass-radius relation by
Vassiliadis \& Wood (1993):

\begin{equation}
\log P=-2.07+1.94\log R_\star-0.9\log m
\end{equation}

\noindent where $R_\star$ is the stellar radius in solar units.

To connect the core mass with the total luminosity of the star, the following 
two equations are used:

\begin{equation}
L=238000\mu_{\rm w}^3Z^{0.04}_{\rm CNO}(m_{\rm c}^2-0.0305m_{\rm c}-0.1802)
\end{equation}

\noindent for stars with core mass in the range $0.5\leq m_{\rm c}\leq 0.66$
(Boothroyd \& Sackmann 1988), and

\begin{equation}
L=122585\mu_{\rm w}^2(m_{\rm c}-0.46)m^{0.19}
\end{equation}

\noindent for stars with $0.95\leq m_{\rm c}$ (Iben \& Truran 1978; Groenewegen
\& de Jong 1993). For stars in the interval $0.66<m_{\rm c}<0.95$, linear
interpolation is performed. In the former equations, $Z_{\rm CNO}$ is the
total abundance (in mass fraction) of carbon, nitrogen and oxygen in the
envelope and $\mu_{\rm w}=4/(5X+3-Z)$
is the average molecular weight for a fully ionized gas, where $X$ and $Z$
are the hydrogen and metal abundances, respectively.

The rate at which the core mass increases as a result of shell H-burning is
computed using

\begin{equation}
\frac{\partial m_{\rm c}}{\partial t}=9.555\times 10^{-12}\frac{L_{\rm H}}{X}
\end{equation}

\noindent where $X$ is the hydrogen abundance (in mass fraction) in the
envelope and $L_{\rm H}$ is the luminosity produced by the H-burning shell (in
solar units). In fact, the latter is computed by the relation (Iben 1977):

\begin{equation}
L=L_{\rm H}+2000(M/7)^{0.4}\exp [3.45(m_{\rm c}-0.96)]
\end{equation}

Finally, the $L-T_{\rm eff}$ relation of the TP-AGB is obtained by
extrapolating the slope of the E-AGB. In fact, to minimize random effects
and possible inconsistencies, the slope is computed using the following
relation, obtained  by bi-logarithmic fit of the slopes of all the models at
the end of the E-AGB:

\begin{equation}
\log (-\frac{\partial L}{\partial T_{\rm eff}})=-2.870+0.994\log L-0.110 \log Z
\end{equation}

\noindent where $Z$ is the metallicity.

Using the former relations, the TP-AGB is computed for each HB and
intermediate age evolutionary track with mass $m\leq 5$M$_\odot$, according
to the intermediate mass limit found by Marigo et al. (1996). The TP-AGB
computation is terminate either when the envelope mass is zero or 
the core mass reaches the Chandrasekhar limit (1.4 M$_\odot$).

\section{Running IAC-star \label{run}}

IAC-star is made available for free use. It can be executed from the
internet site {\tt http://iac-star.iac.es}. A number of interactive
software facilities are intended to be made available from this site,
together with IAC-star itself. In fact, the content and format of the
page is expected to be up-dated, including new software and taking
into account user feed-backs. In the present form, a template is
provided in which several input parameters can be entered. For all
them, default values as well as a brief option list for quick
reference are available. Once all the inputs are provided, the user can
request the program to be executed. This is performed by a dedicated computer at the
Instituto de Astrof\'\i sica de Canarias (IAC). The resulting data
file is stored upon completion in a public access ftp
directory at the IAC. The address of this directory is given to the
user by e-mail, to the address provided by him or herself, together
with the output file name and information on the used inputs and the
output file content. In the following we will describe the input
parameters and the output file content. Since the specific content of both
may change in future versions of the software, only a general description of
them is provided. Detailed, updated information can be found in the IAC-star
internet site {\tt http://iac-star.iac.es}.

\subsection{Input parameters}

The input parameters are introduced as follows (see the IAC-star
internet site, {\tt http://iac-star.iac.es}, for updated details).

\begin{itemize}

\item The chosen stellar evolution and bolometric correction libraries.

\item A seed for the random number generator. The routine used
  for random number generation is ``ran2'', obtained from {\it Numerical
    Recipes in Fortran} (Press et al. 1997).
  
\item Total number of stars computed or saved into the output file. To
  prevent too long runs, the maximum allowed total computed and saved stars
  are limited to some big numbers.
  
\item The minimum stellar luminosity or maximum magnitude in a
  given filter. Although fainter stars are computed, only those brighter than
  this value are saved into the output file. 

\item The SFR, $\psi(t)$. It is computed by interpolation between
  several $\psi(t)$ nodes, defined by the user arbitrarily.
  
\item The chemical enrichment law. Two alternative approaches are allowed to
  produce the chemical enrichment law: (i) just an interpolation between
  several age-metallicity nodes, defined by the user and permitting a
  completely arbitrary metallicity law; (ii) computation from usual
  parameters involved in physical scenarios of chemical evolution, even
  though a self-consistent physical formulation is not intended in this case.
  Metallicity dispersion is allowed in both cases. 
  
\item The IMF. It is assumed to be a power law of the mass, but several mass
  intervals can be defined. 

\item Binary star control. Both the fraction of binary stars and the
  secondary to primary minimum mass ratio can be supplied. 
  
\item Mass loss parameter. Although the default $\eta=0.35$ seems a reasonable
  choice, the user can modify it here. It must be noted that $\eta$
  significantly affects the extension of the HB for low metallicity stars.

\end {itemize}

\subsection{The output file content}

In the current version, the content of the output file is organized as
follows (see the IAC-star internet site, {\tt http://iac-star.iac.es}, for
updated details):

\begin{itemize}
  
\item A head containing information about the input parameters. In
  particular: the libraries used, the total number of computed and stored
  stars; the minimum luminosity or maximum magnitude stored; the current
  SFR, CEL and IMF laws; the fraction of binaries and minimum secondary to
  primary mass ratio, and a heading line for the column content, including a
  list of the photometric bands for which magnitudes have been computed.
  
\item Several lines each containing the information for a single or binary
  star. This information includes physical parameters (luminosity,
  temperature, gravity, mass and metallicity), photometric parameters and
  age. 
  
\item Closing: Integrated quantities are provided in the file closing lines,
  including the total number of ever formed, currently alive and stored
  stars; the total mass ever incorporated into stars (in other words, this is
  just the time integral of the SFR, $\int_0^T \psi(t) {d}t$); the mass
  currently locked into alive stars and into stellar remnants, and the total
  luminosity and the integrated magnitudes. Besides these integrated
  quantities, the logarithm of the sum of squared luminosities
  ($\log\sum_{\rm i} L_{\rm i}^2$, where $L_{\rm i}$ is the luminosity of the
  i-th star if it is single or the total luminosity of the system, if it is
  binary) and the magnitudes derived from this are also given. These are the
  magnitudes associated to surface brightness fluctuations (SBF). In this
  sense, IAC-star can be used as a SBF population synthesis code (see
  Mar\'\i n-Franch \& Aparicio 2004).

\end{itemize}

\section{Final remarks and conclusions \label{conc}}

The program IAC-star, designed to generate synthetic CMDs is presented
in this paper. It calculates full synthetic stellar populations on a
star by star basis, by computing the luminosity, effective temperature
and magnitudes of each star. A variety of stellar evolution and bolometric
correction libraries, SFR, IMF and CEL are allowed
and binary stars can be computed. The program provides also integrated
masses, luminosities and magnitudes as well as SBF luminosity and
magnitudes for the total synthetic stellar population. In this way,
although it is mainly intended for synthetic CMD computation, it can
be also used for traditional and SBF population synthesis research.

Among the main characteristics of the program, the following are the most relevant: 

\begin{itemize}
  
\item Luminosity and effective temperatures of each star are computed by
  direct bi-logarithmic interpolation in the age-metallicity grid of a
  stellar evolution library. Two Padua libraries (Bertelli94 and Girardi00)
  and the new Teramo library (Pietrinferni et al. 2004) are used in the
  current version completed by the Cassisi et al. (2000) models for very low
  mass stars. This produces a smooth distribution of stars in the output
  synthetic CMD which is necessary if accurate studies of SFH are intended.

\item Mass loss is computed during the RGB and the AGB phases.
  
\item The AGB phase is extended to the TP-AGB, covering in this way all the
  significant stellar evolution phases accurately.
  
\item Color and luminosities in a variety of visual broad band, infrared and
  HST filters are provided. Bolometric correction transformations by Girardi
  et al. (2002); Castelli \& Kurucz (2002); Flucks et al. (1994); Lejeune,
  Cuisinier, \& Buser (1997); and Origlia \& Leitherer (2000) are used for
  this purpose.

\end{itemize}

The program is offered for free use and can be executed at the site
{\tt http://iac-star.iac.es}, with the only requirement of
referencing this paper and acknowledging the IAC in any derived
publication. It is intended to produce further improved versions of
the program after feed-back by the user community.

\acknowledgments

We are deeply indebted to Cesare Chiosi, Gianpaolo Bertelli and other
colleagues of the Padua stellar evolution group for a fruitful, long-term
collaboration. Building up IAC-star would not have been possible without this 
collaboration. We are also indebted to Santi Cassisi for
his help in the implementation of the Teramo stellar evolution library
and corresponding bolometric corrections, and for many useful
discussions and advice. The Computer Division of the IAC is
acknowledged for implementing the hardware and the web page interface
of the program. Financial support has be provided by the IAC Research
Division. The authors are funded by the IAC (grant P3/94) and by the
Science and Technology Ministry of the Kingdom of Spain (grant
AYA2001-1661 and AYA2002-01939).

\begin{deluxetable}{lccc}
\tablenum{1}
\footnotesize
\tablewidth{160mm}
\setlength{\tabcolsep}{0.02in}
\tablecaption{Model grids used in the current version of IAC-star
\label{modelos}}
\tablehead{\colhead{Models}  & \colhead{Teramo} & \colhead{Girardi00} & \colhead{Bertelli94}}
\startdata
Mass range [$M_\odot$]& 0.5-10 & 0.15-7 & 0.6-120\\ 
Z range (Z,nstep) &0.0001-0.04 (10) &0.0-0.03 (8) & 0.0004-0.05 (5) \\ 
\tableline
$\alpha=l/H_p$   & 1.913 & 1.68 & 1.63  \\
$Y_{\odot},Z_{\odot}$ & 0.2734, 0.0198 & 0.273, 0.019& 0.282,0.02 \\
Overshooting & canonical \& 0.2H$_p$$^a$ & 0.5H$_p$$^{a,b}$ & 0.5H$_p$$^{a,b}$\\ 
\tableline
EOS            & Irwin04 & Kipp65+Gir96+MHD90 & Kipp65 \\ 
Rad. Opacity   & OPAL96+Alex94 & OPAL92+Alex94 & OPAL92+Hueb77+Cox70\\ 
Nucl. Reactions    & NACRE+Kunz02& Caughlan88+Landr\'e90+WW93& Caughlan88\\ 
\tableline
Model Atm.     & Castelli02& Castelli97& Kurucz92+emp\\ 

\enddata
\tablecomments{Key to references: 
Teramo: Pietrinferni et al. (2004);
Girardi00: Girardi et al. (2000);
Bertelli94: Bertelli et al. (1994);
Irwin04: Irwin, A.W. et al. (2004);
OPAL96: Iglesias \& Rogers (1996);
Alex94: Alexander \& Ferguson (1994);
NACRE: Angulo et al. (1999);
Kunz02: Kunz et al. (2002); 
Castelli02: Castelli \& Kurucz (2002);
Kipp65: Kippenhahn, Thomas \& Weigert (1965);
Gir96: Girardi et al. (1996);
MHD90: Mihalas et al. (1990);
OPAL92: Rogers \& Iglesias (1992);
Caughlan88: Caughlan \& Fowler (1988); 
Landr\'e90: Landr\'e et al. (1990);
WW93: Weaver \& Woosley (1993);
Castelli97:Castelli,  Gratton \& Kurucz (1997);
Huebb77: Huebner  et al. (1997);
Cox70: Cox \& Stewart (1970a,b);
Kurucz92: Kurucz (1992)
} 

\tablenotetext{a}{The prescription on how overshooting is turned off
as a function of mass heavily determines the morphology of the MS-TO
at intermediate ages. Teramo: $\lambda$=(M/M$_\odot$-0.9)/4 for
1.1M$_{\odot}\le$ M$\le 1.7$M$_{\odot}$, $\lambda$=0.2H$_p$ for M$\ge
1.7M_{\odot}$ and $\lambda$=0 for M$\le 1.1M_{\odot}$; Girardi00:
$\lambda=M/M_\odot$-1.0 for 1.0M$_{\odot}\le$ M$\le 1.5$M$_{\odot}$,
with $\lambda$=0.5H$_p$ for M$\ge 1.5M_{\odot}$ and $\lambda$=0 for
M$\le 1.0M_{\odot}$; Bertelli94: $\lambda$=0.5H$_p$ for M$\ge
1.6M_{\odot}$, $\lambda$=0.25H$_p$ for $1.0M_{\odot}\le M \le
1.5M_{\odot}$.}

\tablenotetext{b}{The parameter describing overshooting is its extent
$\lambda$ {\it across} the border of the convective zone, in units of
pressure scale height (H$_p$). This parameter, defined in Bressan et
al. (1981) is not equivalent to others present in the
literature. E.g. $\lambda=0.5$ in the Padova formalism approx
corresponds to 0.25H$_p\ above$ the convective border adopted by the
Teramo group.}

\end{deluxetable}

\newpage
\begin{deluxetable}{lcccccc}
\tablenum{2}
\tablewidth{160mm}
\setlength{\tabcolsep}{0.02in}
\tablecaption{Coeficients of the $\log L - \log m$ and $\log T_{\rm eff} - \log
  m$ fits for very low mass stars
\label{muybaja}}
\tablehead{
\colhead{Relation} & \colhead{$a_0$} & \colhead{$a_1$} & \colhead{$a_2$} &
\colhead{$a_3$} & \colhead{$a_4$} & \colhead{$a_5$}} 
\startdata
Z=0.0002, Y=0.23:                & & & & & & \nl
$\log L$ vs. $\log m$:           & \phn1.1387 & 11.3958 & 15.8615 & \phn8.7825 & \phn0.1773 & \nl
$\log T_{\rm eff}$ vs. $\log m$: & \phn4.2038 & \phn4.3333 & 13.5738 & 22.1148 & 18.2446 & \phn6.0709 \nl

Z=0.0006, Y=0.23:  & & & & & & \nl
$\log L$ vs. $\log m$:       & \phn1.1457 & 11.7429 & 16.6033 & \phn9.4539 & \phn0.3695 & \nl        
$\log T_{\rm eff}$ vs. $\log m$: & \phn4.1851 & \phn4.1282 & 12.2825 & 19.1799 & 15.4070 & \phn5.0853 \nl

Z=0.001,        Y=0.23:     & & & & & & \nl 
$\log L$ vs. $\log m$: &\phn1.0342 & 10.9539 & 14.2405 & \phn6.5840 & --0.8234 & \nl
$\log T_{\rm eff}$ vs. $\log m$: & \phn4.2229 &  \phn4.7253 & 14.9449 & 24.4181 & 20.0889 & \phn6.6370 \nl   

Z=0.002, Y=0.23:                 & & & & & & \nl
$\log L$ vs. $\log m$:       & \phn0.8190 & \phn9.6332 & 10.7591 & \phn2.8464 & --2.2331  &\nl
$\log T_{\rm eff}$ vs. $\log m$: & \phn4.2593 & \phn5.4559 & 18.3152 & 31.0882 & 26.0219 & \phn8.5882 \nl

Z=0.02, Y=0.27:                  & & & & & & \nl 
$\log L$ vs. $\log m$:  & \phn0.3668 & \phn9.8685 & 16.8057 & 14.4196 & \phn3.9769 & \nl
$\log T_{\rm eff}$ vs. $\log m$: & \phn3.8133 & \phn1.5630 & \phn3.2232 & \phn2.8249 & \phn0.7851 & \nl

\enddata
\end{deluxetable}
\newpage

\end{document}